\documentclass[aps,prb,twocolumn,groupeaddress,floatfix,showpax]{revtex4}
\usepackage{graphicx}
\usepackage{afterpage}
\begin{document}
\title{Suppressed reflectivity due to spin-controlled localization in a magnetic semiconductor.}
\author{F.P. Mena$^{1,7}$, J.F. DiTusa$^2$, D. van der Marel$^{3,1}$,
G. Aeppli$^4$, D.P. Young$^2$, A. Damascelli$^5$, J.A. Mydosh$^6$}

\affiliation{$^1$Materials Science Centre, University of
Groningen, The Netherlands}

\affiliation{$^2$Department of Physics and Astronomy, Louisiana
State University, USA}

\affiliation{$^3$D\'epartement de Physique de la Mati\`ere
Condens\'ee, Universit\'e de Gen\`eve, Switzerland}

\affiliation{$^4$London Centre for Nanotechnology, University
College London,UK}

\affiliation{$^5$Department of Physics \& Astronomy, University of
British Columbia, Vancouver, Canada}

\affiliation{$^6$II. Physikalisches Institut, Universit\"at zu
K\"oln, Germany
\\
$^7$Current Address: Space Research Organization of the
Netherlands, Groningen, The Netherlands}
\date{\today}

\begin{abstract}
The narrow gap semiconductor FeSi owes its strong paramagnetism to
electron-correlation effects. Partial Co substitution for Fe
produces a spin-polarized doped semiconductor. The
spin-polarization causes suppression of the metallic reflectivity
and increased scattering of charge carriers, in contrast to what
happens in other magnetic semiconductors, where magnetic order
reduces the scattering. The loss of metallicity continues
progressively even into the fully polarized state, and entails as
much as a 25\% reduction in average mean-free path. We attribute
the observed effect to a deepening of the potential wells
presented by the randomly distributed Co atoms to the majority
spin carriers. This mechanism inverts the sequence of steps for
dealing with disorder and interactions from that in the classic
Al'tshuler Aronov approach - where disorder amplifies the Coulomb
interaction between carriers - in that here, the Coulomb
interaction leads to spin polarization which in turn amplifies the
disorder-induced scattering.
\end{abstract}

\pacs{78.20.-e,71.23.An,71.27.+a,71.30.+h,75.25.+z,78.30.Er}

\maketitle

\section{Introduction}

Future technologies based on the control and state of electron
spins rather than charges are commonly referred to as spintronics.
Efforts to produce materials for spintronics have mostly focused
on thin film III-V semiconductors alloyed with manganese
\cite{ohno92,ohno96,matsukara98,reed01}. In (Ga,Mn)As, the most
fully characterized of these alloys, Mn substitutes a trivalent Ga
ion and acts as a shallow acceptor just above the valence band.
The Mn$^{2+}$ impurities have a local moment associated with a
high spin (S=5/2) configuration, and are ferromagnetically coupled
below the Curie temperature ($T_C$) by a small number of itinerant
hole carriers. In metallic and ferromagnetic (Ga,Mn)As these doped
holes are thought to reside in an itinerant Mn-derived impurity
band \cite{singley02,singley03} about 0.1 eV above the valence
band.

Another route to magnetic semiconductors relies on carrier doping
into narrow band, strongly correlated insulators. Perhaps the most
celebrated is the monosilicide FeSi, which has been investigated
for several decades because it has a large 300 K response to
magnetic fields that vanishes as $T$ approaches zero
\cite{manyala00,manyala04,wernick72,jaccarino67,aeppli92}.
Together with CoSi and the unusual metal MnSi
\cite{mena03,doiron03}, FeSi belongs to the larger group of
transition metal monosilicides, allowing chemical substitutions
across the entire series without change in the cubic B-20 crystal
structure or the nucleation of second phases
\cite{manyala00,wernick72}. Bulk single crystals can be grown and
FeSi can be made metallic and ferromagnetic by the substitution of
Co to form the silicon-based magnetic semiconductor
Fe$_{1-y}$Co${_y}$Si \cite{manyala00,wernick72}. Because the
metals that result are disordered, quantum interference and
electron-electron interactions were found to dominate the
temperature and magnetic field dependent DC carrier
transport\cite{manyala00}. Given that optical properties are a key
to potential spintronic applications \cite{wolf01} and have been
predicted, but not yet observed, to be sensitive to these
interaction effects\cite{altshuler87,lee85,millis84}, we have
measured the optical reflectivity $R(\omega)$ of
Fe$_{1-y}$Co$_y$Si. Here we present the experimental manifestation
of spin-polarization controlled localization on $R(\omega)$ and
the optical conductivity $\sigma(\omega)$ of this magnetic
semiconductor.

\begin{figure}[tb]
 \includegraphics[angle=0,width=8.5cm]{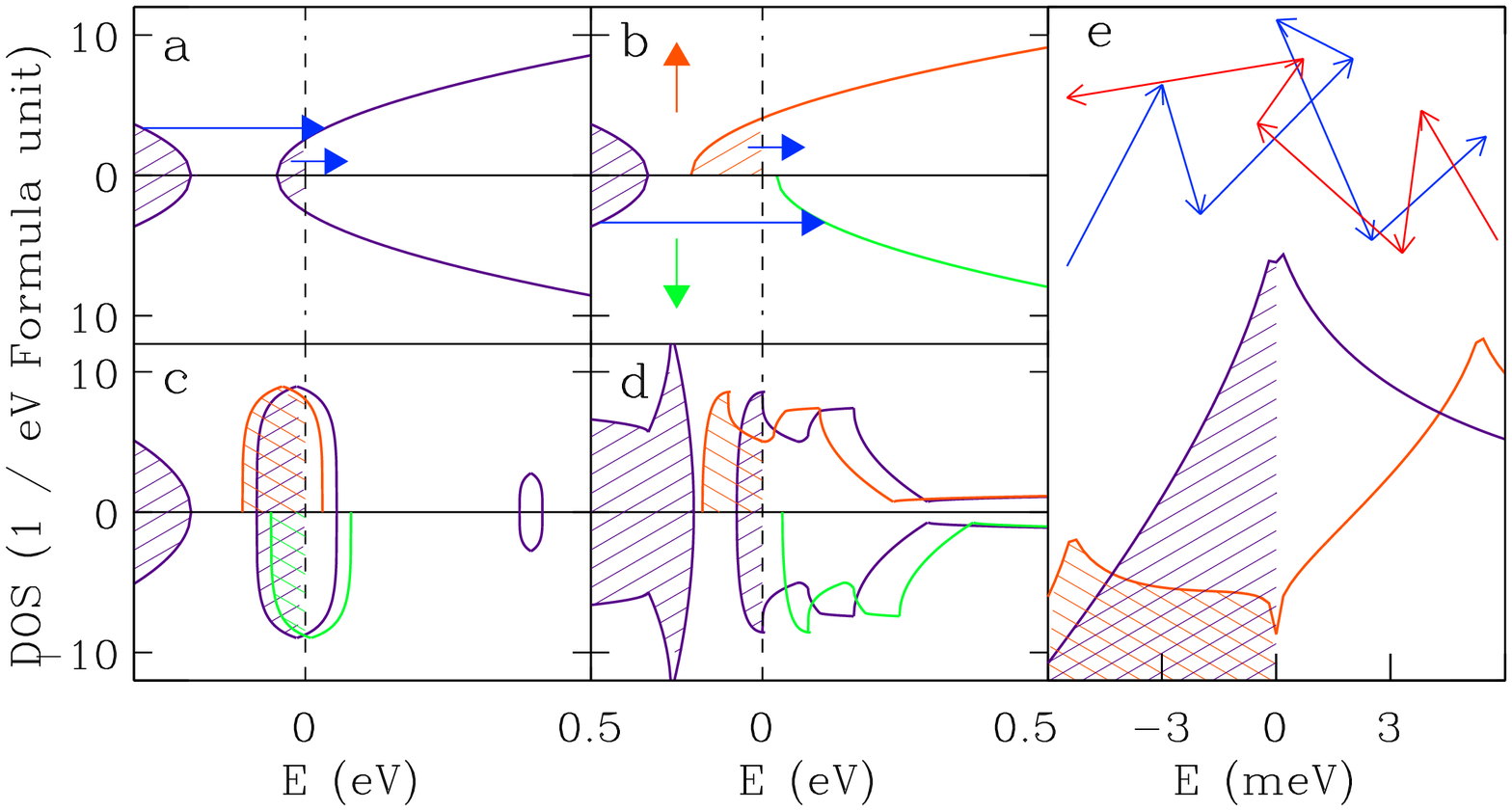}%
 \caption{
Schematic density of states (DOS) for the up and down spin states
of semiconductors, ferromagnetic semiconductors, and ferromagnetic
metals. Blue arrows depict optically allowed interband
transitions: (a) electron doped semiconductor, (b) fully
spin-polarized doped semiconductor, (c) doped semiconductor where
the added carriers reside in an impurity band within the band gap
such as is thought to be the case for (GaMn)As, (d) fully
spinpolarized Fe$_{0.8}$Co$_{0.2}$Si. (e) Disordered conductor
where increased Coulomb interactions lead to an enhanced DOS in
the paramagnetic state (purple curve) and a singular, depleted DOS
at E$_F$ for the spin-polarized state (orange curve). The red and
the blue lines at the top of the figure represent a diffusive path
for two carriers in a disordered metal. Each crossing represents
an interaction event, indicating multiple electron-electron
scattering of the same two carriers without breaking quantum
coherence.} \label{sketch}
\end{figure}
Before presenting the data, we describe what is expected in the
three cases of undoped and doped semiconductors, and the simple
magnetic metals formed from doped semiconductors. For clean
semiconductors, the low-temperature optical conductivity is
dominated by excitations across a gap between valence and
conduction bands. Warming produces holes in the valence band as
well as electrons in the conduction band, adding a Drude-peak
centered at zero energy to the interband transitions. Chemical
doping (Fig. \ref{sketch}a) adds carriers to the valence,
conduction, or impurity bands and yields a zero-temperature
$\sigma(\omega)$ which is very similar to that obtained by warming
in the undoped case - the dominant low-energy feature is a Drude
peak with weight proportional to the carrier density ($n$) and
width $1/\tau$ measuring the scattering rate of the carriers. If
the carriers were spin polarized, either via application of an
external magnetic field or an internal exchange splitting from
spontaneous magnetic order, their band would be split into
majority and minority spin-bands as in Fig. \ref{sketch}b.
Eventually the minority band can be shifted above the Fermi energy
(E$_F$) resulting in a redistribution of all the carriers into the
majority band. The outcome is called a half metallic
ferromagnet\cite{degroot83} (Figs. \ref{sketch}b and
\ref{sketch}d), because electrons carrying the majority spin
belong to a partially filled, metallic band, while those carrying
the minority spin belong to an empty (at $T = 0$) conduction band.
To accommodate all of the itinerant electrons E$_F$ shifts upward,
decreasing the density of states (DOS) at E$_F$ by a factor of
$2^{2/3}$. Naively one might expect this reduction in the DOS to
reduce the low-frequency $\sigma(\omega)$. However, the optical
sum rule states that the integral of $\sigma(\omega)$ over
$\omega$ measures the carrier density $n$, so that conservation of
$n$ in the parabolic band leads to conservation of
$\sigma(\omega)$ through the phase transition as long the
effective mass of the carriers does not change.

This is precisely what has been observed in previous measurements
of spin polarized metals such as CrO$_2$ and NiMnSb, where the
only consequence of the half-metallicity is a slight increase in
the scattering rate at energies above that required to transfer a
carrier between the majority and the unoccupied minority bands
\cite{singley99,mancoff99}. Similarly, measurements of
$\sigma(\omega)$ of the magnetic semiconductor (Ga,Mn)As reveal
changes to the spectrum below $T_C$ due to a reduction of
$\tau^{-1}$ and $m^*$ \cite{singley02,singley03}. As a result
$\sigma(\omega)$ increases at $\omega$ below 350 meV, causing an
increase of the reflectivity, or, in other words, a {\em positive}
magnetic-order-induced reflectivity. Perhaps the most impressive
increases in reflectivity at a ferromagnetic transition occur in
the optical spectra of EuB$_6$  and La$_{1-x}$Sr$_x$MnO$_3$
\cite{degiorgi97,okimoto95}. In these cases cooling into the FM
state is accompanied by large increases in the Drude spectral
weight due to concurrent semimetal- or insulator-metal
transitions.

\section{Experimental Methods}

Single crystals were grown from high purity starting materials
(99.995\% or greater) by either vapor transport, light image
furnace floating zone, or modified tri-arc Czochralski methods.
X-ray spectra showed all samples to be single phase with a lattice
constant linearly dependent on $y$ demonstrating that Co
successfully replaces Fe over the entire concentration range
($0\le y\le 1$). Energy dispersive X-ray microanalysis yielded
results consistent with the nominal concentrations.

The reflectivity of these single crystals was measured from 4 meV
to 0.75 eV while ellipsometry was used to measure directly the
dielectric function from 0.74 to 4.5 eV. In order to monitor in
detail the change of reflectivity as a function of temperature
reported in this manuscript and in a number of other recent
experiments of the same
laboratory\cite{molegraaf02,marel03,kuzmenko03}, typical standard
optical tail cryostats which are commercially available can not be
used. The precision and stability needed in this work requires a
sampling of infrared and optical spectra with a very dense
interval of temperatures, and an overall system stability on the
order of 0.1 \% of the reflected signal or better. We use special
design optical cryostats, which differ from commercial designs in
several important aspects. Further details are described in Ref.
\onlinecite{cryostat_details}.

Between 4 and 750 meV we used Kramers-Kronig relations, along with
a Hagen-Rubens extrapolation of $R(\omega)$ data to $\omega = 0$,
to obtain the phase of $R(\omega)$, and subsequently
$\epsilon(\omega)$. $\sigma(\omega)$ is obtained via the relation
Re$\sigma = (\omega / 4\pi)$Im$\epsilon(\omega)$. We have
carefully checked that $\sigma(\omega)$ is not significantly
altered by our choice of high- and low-$\omega$ terminations. The
Kramers-Kronig output was locked to the ellipsometrically measured
$\epsilon_1(\omega)$ and $\epsilon_2(\omega)$, which strongly
improves the accuracy. The optical reflectivity at low frequencies
and the optical conductivity obtained from the Kramers-Kroning
analysis proved to be consistent with the experimental DC
resistivity.

\section{Optical Reflectivity}
\begin{figure}[tb]
 \includegraphics[angle=270,width=8.5cm]{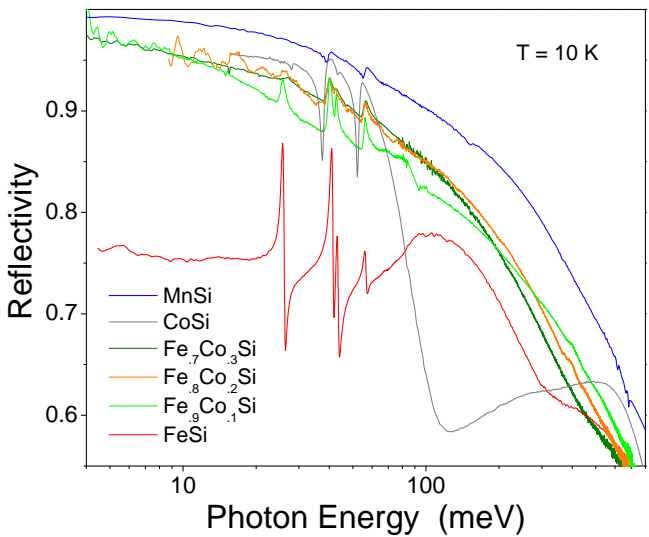}%
 \caption{
Normal incidence reflectivity spectra of FeSi, CoSi,
Fe$_{1-x}$Co$_{x}$Si, and MnSi measured at 10 K. }
\label{all_reflectivity_10k}
\end{figure}
Fe$_{1-x}$Co$_x$Si spans insulating, metallic and polarized
metallic regimes, and is interesting and important because on the
Fe-rich side of the phase diagram\cite{manyala00}, it defies the
standard expectations for all three. In Fig.
\ref{all_reflectivity_10k} we compare the reflectivity spectra at
10 K of FeSi,CoSi, MnSi and Fe$_{1-x}$Co$_x$Si with x=0.1, 0.2,
0.3. The latter three samples order magnetically at $T_C$ = 10, 36
and 49 K, while MnSi orders at 29.5 K\cite{manyala00,wernick66}.
All samples shown except FeSi become 100 \% reflecting in the
limit $\omega\rightarrow 0$, as expected because they are good
electrically conducting materials. The reflectivity spectra reveal
optical phonons at approximately 25, 40, 43 and 56 meV, the shape
of which vary progressively from the asymmetric dispersive variety
for the semiconducting samples to negative dips for the more
metallic samples. This is a direct manifestation of the negative
term contributed to the dielectric function by the free charge
carriers, in addition to the optical phonon oscillators. MnSi is a
relatively good metal at these temperatures, causing the strong
screening of the optical phonons in the reflectivity spectra. In
an earlier publication the infrared reflectivity of MnSi has been
reported to increase when it becomes magnetically
ordered\cite{mena03}. This is a natural consequence of the
suppression of the spin-disorder scattering of the electrons,
simultaneously suppressing the scattering rate of the free charge
carriers and increasing the amplitude of the low frequency
dielectric constant.

\begin{figure}[tb]
 \includegraphics[angle=270,width=8.5cm]{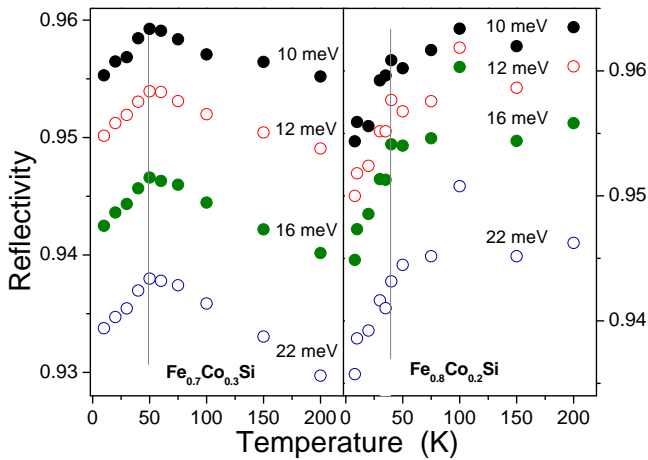}%
 \caption{
Temperature dependence of the reflection coefficient of
Fe$_{0.8}$Co$_{0.2}$Si and Fe$_{0.7}$Co$_{0.3}$Si for four
different infrared photon energies. The vertical gray lines
indicate the corresponding magnetic ordering temperature.}
\label{reflectivity_T_dependence}
\end{figure}
The central experimental result of this paper is the temperature
dependence of the reflectivity of Fe$_{1-x}$Co$_{x}$Si, shown in
Fig. \ref{reflectivity_T_dependence}.
The different behaviour of the paramagnetic state also exists for
the transport properties, and it reflects the closer proximity to
the metal insulator transition of samples with a smaller carrier
concentration. Both Fe$_{0.8}$Co$_{0.2}$Si and
Fe$_{0.7}$Co$_{0.3}$Si behave exactly opposite to MnSi when the
magnetic order occurs: The reflectivity of both samples is clearly
{\em suppressed} in the ordered state, {\em i.e.} in the
magnetically ordered phase the material becomes less metallic as
compared to the paramagnetic phase.
We observed no spin-ordering induced change of reflectivity in the
10 \% doped sample, which is not surprising: $T_C$ of this sample
is 10 K. Assuming that the spin-ordering induced changes are
similar to those of the other two samples, the estimated change of
reflectivity between 6 Kelvin (the lower limit of this cryostat)
and 10 Kelvin is only about 0.0005, which is too small to detect
with the current accuracy.

\section{Optical Conductivity}

\begin{figure}[tb]
 \includegraphics[angle=0,width=8.5cm]{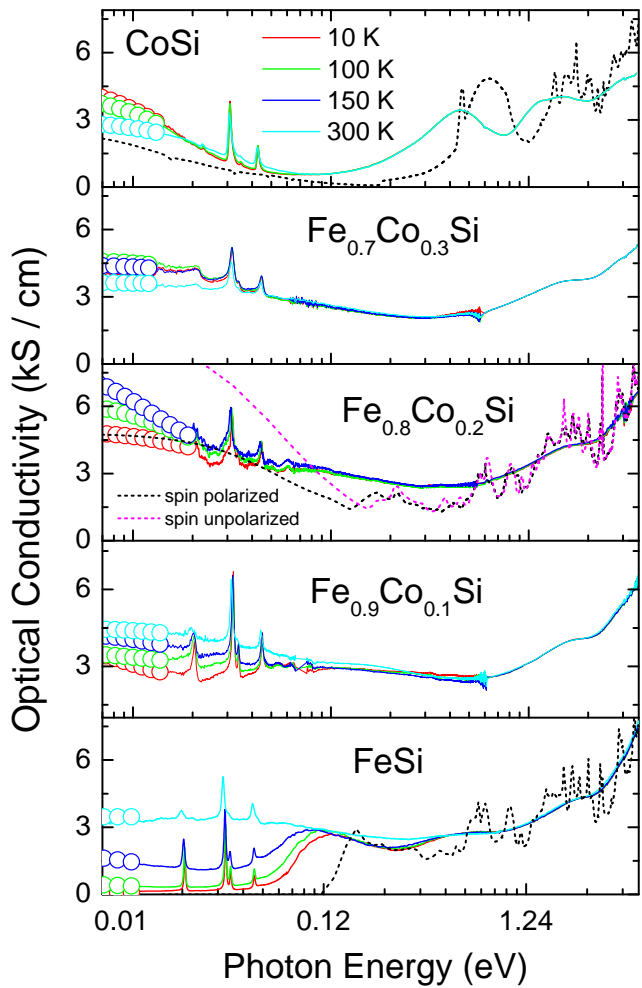}%
 \caption{
Experimental optical conductivity spectra of Fe$_{1-x}$Co$_{x}$Si
for x=0.0, 0.1, 0.2, 0.3 and 1.0. (solid lines and open circles).
The calculated optical conductivity using the Local Density
Approximation is shown for FeSi, Fe$_{0.8}$Co$_{0.2}$Si and CoSi
(olive dashed cures). Widths ($\tau^{-1}$) of the Drude peaks were
chosen to minimize the deviation from the experiment spectra. For
Fe$_{0.8}$Co$_{0.2}$Si, the calculation corresponding to full
spin-polarization (purple dashed curves) is a much better estimate
for the Drude weight than unpolarized calculation (olive dashed
curve), even at 300 K, consistent with short range magnetic order
well above $T_C$.}
 \label{conductivity}
\end{figure}
In order to obtain a clear experimental picture of this effect, we
examine the optical conductivity spectra for a few selected
temperatures, shown in Fig. \ref{conductivity} for the
semiconducting parent compound FeSi, and Fe$_{1-x}$Co$_{x}$Si. We
begin with pure CoSi (top frames in Fig.\ref{conductivity}), long
known as a diamagnetic metal with a very low carrier density,
$\sim 1\%$ of electrons/formula unit \cite{asanabe64}. Our data
agree with the simple ideas outlined above; {\em i.e.} there is a
small Drude peak centered at $\omega = 0$ which coexists with
interband transitions beginning at $\sim 125$ meV. At $T = 10$ K,
the width of the Drude peak is $\hbar/\tau$= 20 meV, which in the
simplest analysis implies a carrier mean free path of 5 nm. The
main effect of warming to 300 K is to broaden the Drude peak by an
amount less than $k_BT$.

The lower frame of Fig. \ref{conductivity} displays data for the
alloy's other end member, insulating FeSi. The optical
conductivity, which shows no hint of a band gap at 300 K, is
almost completely suppressed below 75 meV upon cooling to 10 K, as
in Ref. \onlinecite{schlesinger93} which emphasized that in a
conventional band picture, remnants of a 60 meV gap should be
clearly visible at 300 K (26 meV). This disappearance of the
semiconductor gap at approximately 200K has been confirmed by
subsequent investigations of the optical
properties\cite{degiorgi94,damascelli97}. A second important
feature is that the energy range over which $\sigma(\omega)$
changes as a function of temperature, is extremely large as
indicated by the logarithmic $\omega$ scale in Fig.
\ref{conductivity}. Careful analysis of the temperature dependence
of the integrated spectral weight\cite{mena04} confirms the result
of Ref. \onlinecite{schlesinger93}, that there is a lack of
spectral weight conservation to energies above 80 times $E_g$.
Even if the spectral weight would be distributed over a wide
spectral range in an unmeasurable way, it is not recovered on the
energy scale of a few times the gap as expected.

While the obliteration of all remnants of a 60 meV gap at 300 K
can be attributed, at least in part, to the feedback of
vibrational disorder on the electronic structure using a
conventional band picture where electron-electron correlations are
treated within the framework of the Local Density
Approximation\cite{jarlborg99}, the issue of the spectral weight
redistribution appears to require a more rigorous treatment of the
correlation effects. Urasaki and Saso\cite{urasaki99} treated the
bands on either side of the gap with the 2-band Hubbard model,
while assuming a moderate $U$ ($U=0.5$ eV). Their calculated
optical spectra agree in detail with the experiments. In
particular their model explains both the temperature shift of gap
edge, and the temperature dependent spectral weight
redistribution. Taken together this suggests the important role of
electron-correlations for the physical properties of FeSi
\cite{urasaki98,urasaki99,rozenberg96,fu94}.

Doping FeSi via substitution of Co for Fe yields similar problems
for the standard model underlying semiconductor optics. The
conductivity of the 10 percent doped sample exhibits a suppression
below 10 meV for T $<$ 50 K, but there is no complete gap like in
FeSi at the same temperatures. In contrast to the data for pure
CoSi, but in agreement with our findings for FeSi at 300 K,
Fe$_{1-x}$Co$_{x}$Si for the three dopings considered here
displays a $\sigma(\omega)$ which decays weakly from $\hbar\omega
= 0$ to 350 meV, while all traces of the gap in the pure FeSi
parent are obliterated \cite{chernikov97}. Even with the
assumption of a scattering rate in excess of the 60 meV gap of
FeSi, a simple Drude analysis (dashed lines in
Fig.\ref{conductivity}) based on our band structure calculations
cannot account for $\sigma(\omega)$. We conclude that treating the
electrons in Fe$_{1-y}$Co$_y$Si as a simple Fermi liquid formed in
the conduction band is incorrect, notwithstanding the remarkable
simplicity of aggregate $\omega = 0$ properties such as the normal
Hall effect and ordered magnetization, which correspond to one
carrier and one polarized spin per Co atom \cite{manyala00}.

Beyond showing that the parent insulator and its electron-doped
derivative violate standard ideas about undoped and doped
semiconductors, Fig.\ref{conductivity} also reveals that
Fe$_{1-y}$Co$_y$Si defies expectations for itinerant magnets. In
particular, cooling yields a suppression of the low-frequency
conductivity of a different qualitative nature than seen for FeSi,
where it occurs throughout the gapped region. The suppression of
the optical conductivity is of course consistent with the
suppression of the dc conductivity from the dc transport
measurements, as well as the raw reflectivity data $R(\omega)$ of
Fig. \ref{reflectivity_T_dependence}. Thus, in contrast to what
occurs for all other metallic ferromagnets, including MnSi
\cite{mena04} and (Ga,Mn)As \cite{singley02,singley03}, the
approach and onset of magnetic order at $T_C = 36$ K decreases the
conductivity and suppresses the metallic screening of
Fe$_{1-y}$Co$_y$Si.

\begin{figure}[tb]
 \includegraphics[angle=0,width=8.5cm]{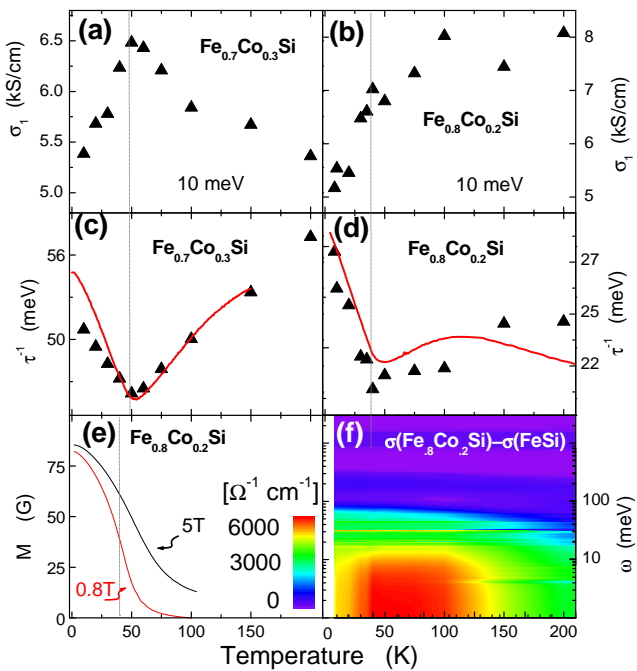}%
 \caption{
Temperature dependence of related physical properties of
Fe$_{1-x}$Co$_{x}$Si for two different doping concentrations. The
optical conductivity $\sigma(\omega)$ for $\hbar\omega = 10$ meV
({\bf a} and {\bf b}), the corresponding optical scattering rates
$\hbar/\tau$ (triangles in {\bf c} and {\bf d}), DC resistivities
scaled as to overlay the scattering rates (red solid curves in
{\bf c} and {\bf d}), bulk magnetization of Fe$_{0.8}$Co$_{0.2}$Si
({\bf e}), and optical conductivity difference of
Fe$_{0.8}$Co$_{0.2}$Si and FeSi for a large range of temperatures
and frequencies {\bf f}. The corresponding magnetic ordering
temperatures are indicated by the grey lines. }
 \label{all_T_dependent_properties}
\end{figure}
Figure \ref{all_T_dependent_properties} reveals more detail on the
evolution of the optical data with $T$, and compares them to
transport and magnetization data. $R(\omega)$ in
Fig.\ref{reflectivity_T_dependence} at low $\omega$ simply follows
the $\sigma(\omega=0)$ (Fig. \ref{all_T_dependent_properties}),
which experiences its main drop below $T_C$. For higher $\omega$
(not shown), $R(\omega)$ decreases continuously from 300 K, with
no visible anomaly at $T_C$ for photon energies larger than 40
meV. The obliteration of singular behaviour near $T_C$ with
increasing $\omega$ is in accord with the extended critical
regime, or superparamagnetism (field induced short range order),
indicated by the magnetization data of
Fig.\ref{all_T_dependent_properties}d. Here a modest (compared to
$k_BT$) external field of 5 T produces very appreciable
polarization to $T$'s as high as 100 K$\sim 3T_C$.
Fig.\ref{all_T_dependent_properties}f shows $\sigma(\omega,T)$ of
Fe$_{0.8}$Co$_{0.2}$Si after subtraction of $\sigma(\omega)$ of
FeSi, revealing where the added carriers reside in the excitation
spectrum of the nominally pure compound. Cooling builds up the
differential (relative to the insulator) spectral weight down to
$T_C$, whereupon there is a loss especially apparent below 25 meV.
Again, our observation of a reduced spectral weight below $T_C$
contradicts both the standard models based on independent
quasiparticles and measurements for other magnetic semiconductors
\cite{singley02,singley03}.

In addition, cooling below $T_C$ causes the wide Drude-like peak
to broaden, rather than to narrow.  We have parameterized the
low-$\omega$  $\sigma(\omega)$ by fitting the standard Drude form,
superposed on a $T$-independent Lorentzian peak at 800 cm$^{-1}$,
to the data. Fig.\ref{all_T_dependent_properties} displays the
resulting scattering rates ($\tau^{-1}$), which track the bulk
resistivities ($1/\sigma_{1}(\omega=0)$) and undergo a sharp
upturn below $T_C$. The important and unique contribution of the
optical data is to show that the unusual rise in resistivity below
$T_C$ is due to enhanced scattering as opposed to a reduction of
the charge carrier density/mass ratio.

\section{Discussion}
\begin{figure*}[tb]
 \includegraphics[angle=90,width=17cm]{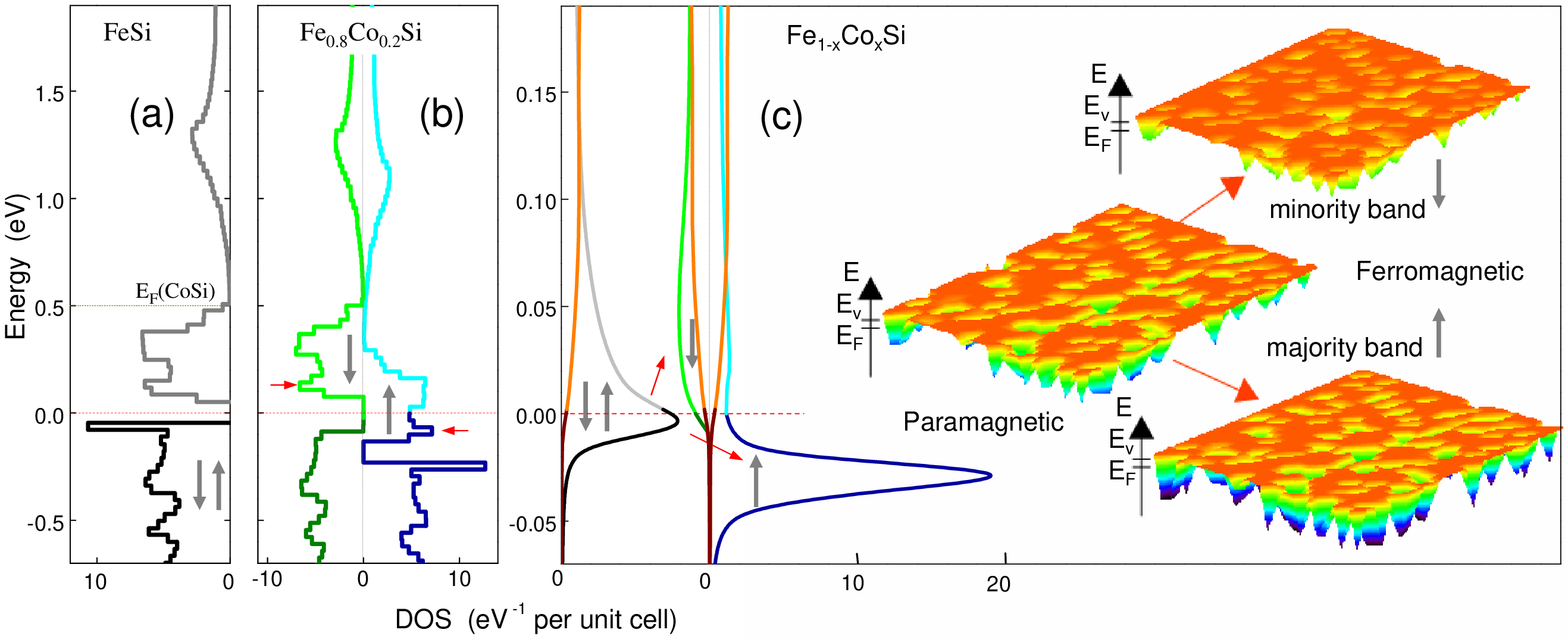}%
 \caption{
(a) Calculated density of states of FeSi. The Fermi energy (red
line) is inside the semiconductor gap. The band-calculations of
CoSi give a very similar DOS, with the Fermi energy shifted to a
higher energy (dark dotted line). (b) DOS of
Fe$_{0.8}$Co$_{0.2}$Si using the virtual crystal
approximation\cite{guevara04}. These results confirm the full spin
polarization of Fe$_{0.8}$Co$_{0.2}$Si. Arrows indicate $E_F$ in
unpolarized state. (c) Local DOS at the Co atom substituting Fe in
FeSi in the unpolarized state (grey curves) and the spin-polarized
state (blue and green curves), calculated using the tight-binding
Green's function formalism\cite{wolff61,clogston62}, while
treating the on-site Hubbard repulsion in the mean field
approximation\cite{anderson61,anderson78}. The orange curves are
the DOS of the FeSi host material. The Fermi energy ($E_F$) lies
just above the FeSi conduction band edge ($E_c$). The potential
energy landscape for the carriers due to the Co impurities is
sketched on the righthand side. In the ferromagnetic state
carriers in the majority spin sub-band scatter more strongly from
the random potential than either carriers in the minority spin
sub-band or in paramagnetic bands due to the increased depth of
the potential wells. }
 \label{dosses}
\end{figure*}
To begin to understand our data, we have calculated the density of
states (Fig.\ref{dosses}) as well as $\sigma(\omega)$
(Fig.\ref{conductivity}) derived from the band structure of
Fe$_{1-y}$Co$_y$Si. These calculations employ the Local Density
Approximation (LDA) with the Gunnarsson-Lundquist exchange
correlation potential carried out self consistently using the full
potential linear muffin-tin orbital
method\cite{gunnarsson76,savrasov92,mazin88}. For
Fe$_{0.8}$Co$_{0.2}$Si a non-integer charge to iron in FeSi was
assigned. We evaluated $\sigma(\omega)$ with a k-space integration
over 216 points in the irreducible part of the Brillouin
zone\cite{mena04}. The band structure agrees with the
magnetization and Hall effect and more complicated supercell
calculations, in that the minority spin Fermi surface resides in a
gap \cite{manyala00,manyala04,guevara04}. It also supports the
picture that Co doping merely adds electrons to bands inherited
from the FeSi parent, revealing, as does experiment, that CoSi
(with one extra electron/Fe) is a very low $n$ metal.

While treating electron-correlations within the framework of the
LDA (Fig.\ref{dosses}) has some successes (for example in
explaining the ground state properties), for the description of
the relevant (magneto-) transport and optical properties it is
necessary to take into account the impact of electronic
correlations on the excited states. Generally speaking this
requires a treatment beyond the Local Density Approximation.
Important experimental observations which need to be understood
are the apparent loss of carriers at low-$\omega$, and the
increase--instead of the conventional decrease--in $\tau^{-1}$
below $T_C$ of Fe$_{1-y}$Co$_y$Si. To make progress, the Coulomb
interactions need to be considered. How these underpin the moment
formation at modest $T$, as well as the rapid filling of the gap
in the parent compound via the paradigm of the Kondo insulator, is
discussed elsewhere \cite{urasaki98,urasaki99,rozenberg96,fu94}.

Why do we observe an increased scattering rate below $T_C$,
opposite in sign to Mn doped GaAs and MnSi? For weak
spin-polarization, this effect most likely follows from the
interplay of Coulomb coupling and the quantum mechanical
interference of diffusing charge carriers\cite{altshuler87,lee85}.
This feature of disordered metals, hitherto seen only in transport
and tunneling, but witnessed optically for the first time in our
experiments, is demonstrated schematically in Fig.\ref{sketch}g.
Arrows at the top of the figure demonstrate diffusive paths for
two carriers in a disordered metal. The large elastic scattering
rate results in multiple scattering of these same two carriers
without breaking quantum coherence, effectively enhancing their
Coulomb interaction because the carriers are visible to each other
more often. The increased Coulomb coupling induces square-root
singularities in the density of states at
$E_F$\cite{altshuler87,lee85}. Spin-polarization, either from
external magnetic fields or a spontaneous magnetization, shifts
the singularities with respect to $E_F$ resulting in a reduction
of $\sigma(\omega)$\cite{altshuler87,lee85}, illustrated
schematically in the figure. Thus $\sigma(\omega)$ and $R(\omega)$
display singular behaviour in $T$ at low-$\omega$ just as we
observe. MnSi, the only other known magnet in the transition metal
monosilicide group, is not a disordered metal, in accordance with
our observation of no magnetic-order-induced reduction of the
reflectivity\cite{mena04}. However, well below $T_C$ the
saturation magnetization of Fe$_{1-x}$Co$_x$Si with $x < 0.3$
approaches\cite{manyala00} that of a halfmetallic ferromagnet
(Fig. \ref{sketch}d). In this limit one of two spin components is
absent near the Fermi energy, a state of affairs quite distant
from the situation considered in Fig.\ref{sketch}g. As we will
explain below, disorder and Coulomb interactions appear to control
the optical and magnetotransport anomalies of these compounds {\em
also} in the limit of strong spin-polarization.

The explanation here stems from the strong 3d-character of the
conduction bands. In a tight binding picture of Fe$_{1-x}$Co$_x$Si
the random substitution of Co for Fe within the FeSi lattice
corresponds to deepening the potential-well of the Co-sites,
allowing one extra electron per site to enter. This attractive
potential causes a downward shift of the local DOS of the Co-atoms
relative to the FeSi DOS. Provided that this potential exceeds a
critical value, an impurity state will be pulled below the
conduction band, {\em i.e.} inside the semiconductor gap. However,
even if the potential is smaller than the critical value, the
local DOS on the Co-sites has a pile-up of states near the gap
(grey curves in Fig. \ref{dosses}c). In this scenario the elastic
scattering of the conduction electrons is a consequence of the
disordered potential landscape (Fig. \ref{dosses}c) and, in the
case of Fe$_{1-x}$Co$_x$Si, brings the system close to an Anderson
metal-insulator transition. How close is indicated by the product
$k_Fl$ of the Fermi vector and mean free path which we measure to
be 6.9 if we assume a single Fermi surface, and 1.3 if we consider
that there are 12 electron pockets \cite{hamann93}. As this weakly
metallic electron gas is cooled, the spin splitting of the bands
due to the exchange interaction shifts the majority and minority
bands of Co-3d character with respect to the FeSi conduction band.
The consequence for the disordered potential landscape is
illustrated in the right hand side of Fig. \ref{dosses}c where the
depth of the potential wells in proximity to the Co-sites is
increased (decreased) for the majority (minority) spin sub-band.
Consequently, the majority spins are more strongly scattered
compared to the paramagnetic, or unpolarized high temperature,
state. The net magnetoresistance of two parallel channels is
\cite{fert68}
$ \rho^{-1}=\rho_{\uparrow}^{-1}+\rho_{\downarrow}^{-1}$
where $\rho_{\sigma}$ are the resitivities of each spin-channel
separately, if spin flip scattering, such as arising from
spin-orbit coupling, can be neglected. In the limit of vanishing
spin-polarization the sign of the magnetoresistance depends on the
microscopic details of the model describing $\rho_{\sigma}$ as a
function of doping\cite{fert68}. However, in the present case an
almost fully spin-polarized state is observed at low temperatures,
hence we consider here the limit where the spin-splitting of the
energy levels is large. Due to hopping between the Co-sites the
states inside the gap form bands, and at the high doping levels
considered here the width of these bands is a considerable
fraction of the bandwidth of FeSi. For moderate spin-splitting,
these bands merge with the FeSi conduction band, and the DOS at
$E_F$ remains ungapped, and the density of states still resembles
that given  by the virtual crystal approximation\cite{guevara04}
presented in Fig. \ref{dosses}b. Ultimately, for a sufficiently
large spin-splitting the majority spin-states become separated
from the conduction band by a gap. Since the number of doped
electrons is exactly equal to the number of majority spin-states,
the Fermi-level resides inside this gap. Consequently this state
of matter is not only fully spin-polarized, but it is also an
insulator. In other words, in the limit of large spin-polarization
both $\rho_{\uparrow}$ and $\rho_{\downarrow}$ diverge, the former
because all localized up-states are occupied, the latter because
the down-states are empty in this limit. While these extreme
conditions appear not to be met in the case of Co-doped FeSi, the
majority spins in this material do outnumber the minority spins by
a large fraction below $T_C$, hence the dominant effect at high
spin-polarization is a positive magneto-resistance, just as we
observe in Fig. \ref{all_T_dependent_properties}.

The Al'tshuler-Aronov description and the model outlined above
both take elastic scattering and Coulomb interactions as the
starting point, and both lead to the same conclusion, namely that
spin-polarization of a disordered metal leads to enhanced
scattering. Whereas the former captures the subtle low energy
scale effects caused by the coherent scattering between
impurities, the latter becomes relevant in the limit of strong
local perturbing potentials and a high degree of
spin-polarization. Actually, in our description, we consider first
the effect of the Coulomb interaction as the cause of spin order
and therefore, via exchange splitting of the bands, an amplifier
of potential scattering,  whereas Al'tshuler-Aronov considers
disorder as source of scattering which then amplifies the Coulomb
interaction. The unification of these two limiting cases in a
single framework remains a theoretical challenge yet to be met,
although we suspect that a two-parameter scaling description might
eventually account for the remarkable success of the
Al'tshuler-Aronov expressions in accounting for the
magnetoconductance of FeCoSi over such a wide range of fields and
temperature\cite{manyala00}.

The reason for the differences from $\sigma(\omega)$ of (GaMn)As
now becomes clear; the bands with Mn d-character are well below
the GaAs valence band and produce a local magnetic moment
interacting with the conducting holes. Therefore the effect of
polarization on the carrier elastic scattering is less important
and only small changes to $\sigma(\omega)$ due to variations in
the spin disorder scattering are observed.

\section{Conclusions}

We have shown that the optical properties of Fe$_{1-y}$Co$_y$Si
are very different from those for (Ga,Mn)As, even though bulk
properties such as the off-diagonal conductivity are remarkably
similar\cite{manyala00,manyala04}. Doping produces an optical
response throughout the gap of FeSi, implying that we are dealing
not with an impurity band as in (Ga,Mn)As---but rather with
carriers donated to the conduction band of FeSi. Finally, we have
discovered an optical reflectivity which decreases rather than
increases upon entering the spin-polarized state. The
corresponding rise in the scattering rate, uniquely visible in the
optical data, demonstrates that the origin of this effect is the
Coulomb interaction between electrons in a disordered system.

\section{Acknowledgements} This work was supported by the National
Science Foundation under contract No. DMR0406140, a Wolfson-Royal
Society Research Merit Award, the Basic Technologies program of
the UK Research Councils, the Swiss National Science Foundation
through the NCCR 'Materials with Novel Electronic Properties', and
the Netherlands Foundation for Fundamental Research on Matter with
financial aid from the Nederlandse Organisatie voor
Wetenschappelijk Onderzoek. We gratefully acknowledge A. A.
Menovsky, C. Presura and A.I. Poteryaev for their assistance with
crystal growth, optical experiments and LDA calculations
respectively.

\end{document}